\documentclass{SCIS2021}
\usepackage{multirow}
\usepackage[normalem]{ulem}
\useunder{\uline}{\ul}{}
\usepackage{ragged2e}
\newcolumntype{Y}{>{\centering\arraybackslash}X}

\begin{document}
\ArticleType{REVIEW PAPER}
\Year{2020}
\Month{}
\Vol{}
\No{}
\DOI{}
\ArtNo{}
\ReceiveDate{}
\ReviseDate{}
\AcceptDate{}
\OnlineDate{}

\title{A Review of Machine Learning-based Failure Management in Optical Networks}{Machine learning for failure management}

\author[1]{Danshi WANG}{}
\author[1]{Chunyu ZHANG}{}
\author[1]{Wenbin CHEN}{}
\author[1]{Hui YANG}{}
\author[1]{\\Min ZHANG}{{mzhang@bupt.edu.cn}}
\author[2,3]{Alan Pak Tao LAU}{{alan.pt.lau@polyu.edu.hk}}

\AuthorMark{Wang D S}

\AuthorCitation{Danshi Wang, et al}


\address[1]{ State Key Laboratory of Information Photonics and Optical Communications, \\Beijing University of Posts and Telecommunications (BUPT), Beijing {\rm 100876}, China}
\address[2]{Photonics Research Center, Department of Electrical Engineering, \\The Hong Kong Polytechnic University, Hong Kong SAR, China}
\address[3]{The Hong Kong Polytechnic University Shenzhen Research Institute, Shenzhen {\rm 518057}, China}

\abstract{Failure management plays a significant role in optical networks. It ensures secure operation, mitigates potential risks, and executes proactive protection. Machine learning (ML) is considered to be an extremely powerful technique for performing comprehensive data analysis and complex network management and is widely utilized for failure management in optical networks to revolutionize the conventional manual methods. In this study, the background of failure management is introduced, where typical failure tasks, physical objects, ML algorithms, data source, and extracted information are illustrated in detail. An overview of the applications of ML in failure management is provided in terms of alarm analysis, failure prediction, failure detection, failure localization, and failure identification. Finally, the future directions on ML for failure management are discussed from the perspective of data, model, task, and emerging techniques.}


\maketitle

\section{Introduction}
In the present big data era, data is growing exponentially, and optical networks serve as the backbone for high-capacity and long-distance data transmission. The optical networks are always large scale, comprise massive components, and cover a wide area. Considering this, instances of failure will cause extremely serious consequences, such as massive data loss, large-scale computing interruption, core information transfer blocking. All users including the government, private enterprises, financing institutions, transportation industry, manufacturing industry, and individuals would suffer heavy economic losses. Therefore, failure management in optical networks is crucial to ensure the stable operation, maintain the service status, and, in the event of a failure, recovery the failure rapidly.

Optical networks are subject to several types of failure, primarily divided into soft and hard failure. These typically include fiber cut, filter effect, laser drift, component (e.g., optical module, optical amplifier, optical switch) breakdown, and system aging. Handling network failures can be accomplished at different levels, i.e., alarm analysis, failure prediction, failure detection, failure identification, failure magnitude estimation, and failure localization ~\cite{musumeci2019tutorial}. Conventional failure management methods typically implement these functions using simplified threshold methods or probability statistics models. However, these are effective only for simple and static cases. For complex and dynamics cases, expert manual intervention is still required, which leads to high labor costs and inevitable personal errors.  

Recently, techniques from artificial intelligence (AI) have been widely studied to address multiple problems in optical networks, such as traffic prediction, topology design, path computation, resource allocation, as well as failure management ~\cite{rafique2018machine,musumeci2018overview,khan2019optical}. From conventional machine learning (ML) to neural network-based deep learning (DL), various algorithms in AI communities can help analyze and process a large amount of data and information collected from the continuous activity of a huge number of monitors and alarms. Based on the available data and objectives of a given model, the proper learning algorithms are selected and modified to fulfill the corresponding tasks in terms of different failure scenarios ~\cite{musumeci2021machine}. In this manner, a variety of intelligent, accurate, and low-cost solutions are developed. 

As a tutorial paper, literature~\cite{musumeci2019tutorial} presented a gentle introduction to ML-based failure management and provided the guidance on how to explore the following researches about this topic, but not a comprehensive review article. Even several ML-related survey or review papers have been published, literatures~\cite{rafique2018machine,musumeci2018overview,khan2019optical} involve the various applications of ML on all sides of optical communications and networks, but not specialized on failure management. All of these papers were published in its infancy before 2019. However, as we known, applications of ML in failure management grows rapidly during recent years, but still lacking of a complete survey on this special topic that summarizes the up-to-date works.

In this study, we review the applications of ML to failure management in optical networks from infancy to the near term. First, we introduce the background of failure management and interpret the typical tasks. The key physical objects that need health monitoring in optical networks and the potential failure categories for each object are listed in detail. Then various ML algorithms applied for failure management are depicted. ML algorithms are strongly dependent on data, and thus, the data sources with data content and extracted information in optical networks are also discussed. Following that, we survey the existing schemes of ML-based failure management in terms of alarm analysis, failure prediction, failure detection, failure localization, and failure identification. Finally, the future scope of this topic is envisioned from the perspective of data, model, task, and emerging techniques.

\section{Background on Failure Management in Optical Networks}
\subsection{Concept of Failure Management}
The objective of failure management is to detect, isolate, and repair all types of network faults, ensure the reliable and stable operation of the network, and meet the customer's service level agreement. Tasks in failure management are generally divided into active and passive methods, which can be roughly divided into alarm analysis, failure prediction, failure detection, failure identification, failure diagnosis, failure localization~\cite{wang2021machine}, as illustrated in Fig. 1.  

\begin{figure}[htbp]
\centering
\includegraphics[width=12cm]{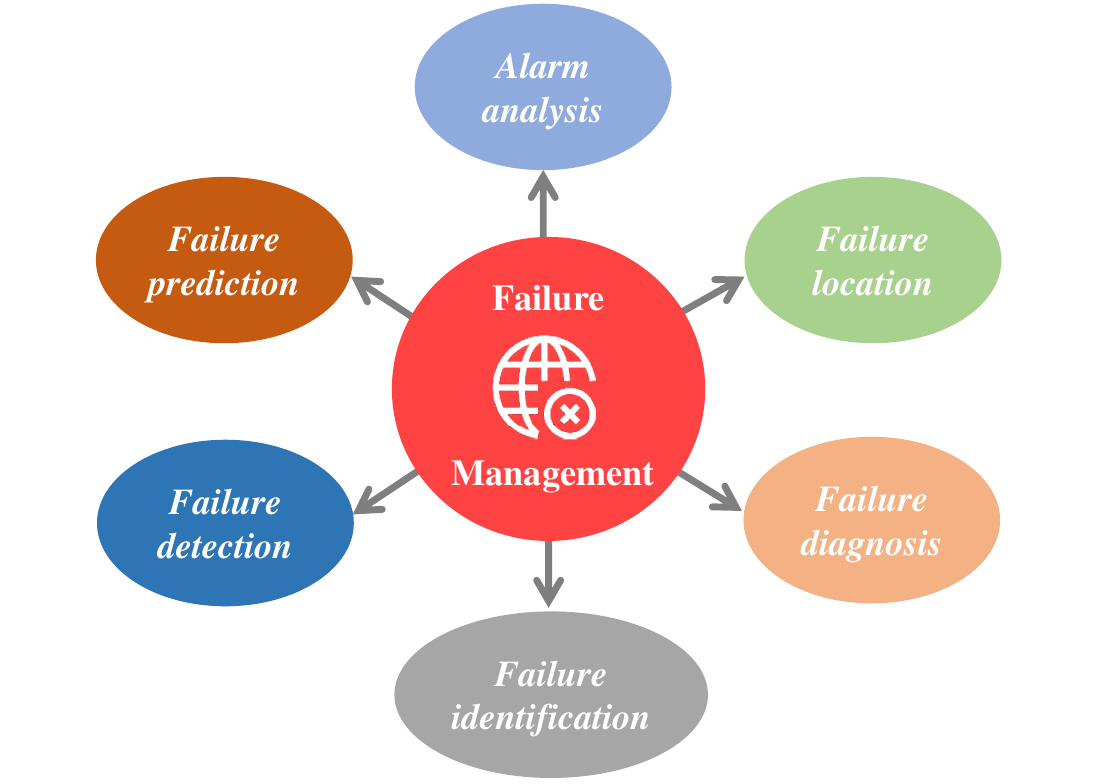}
\caption{Concept of failure management consisting of alarm analysis, failure prediction, failure detection, failure identification, failure diagnosis, failure localization, etc.  }
\label{fig1}
\end{figure}

The alarm analysis conducts fault management on the monitoring object by studying the alarm data and information. When a fault occurs, massive alarms with different types and levels are often accompanied by network management. The current alarm analysis is primarily divided into two categories: one is alarm prediction with the purpose of predicting whether this type of alarm will occur in the future; the other is relationship analysis between alarm data using data mining or ML methods to find the root cause and locate the fault according to the localization information of the root cause.

Failure prediction is a proactive defense method that predicts the possibility of equipment abnormality through certain signs (generally received from various monitors) before a failure occurs, and takes active measures to prevent failure. Failure detection, identification, diagnosis, and localization are passive methods that are a series of actions taken after a failure occurs, with the aim of recovering the fault system as soon as possible. Failure detection focuses on confirming whether a fault occurs after the fault warning or alarm, and its goal is to quickly detect the failure severity (i.e., health or different fault degrees). 

After fault detection, identifying the fault type is crucial for recovering the faulty system. Failure identification is typically modeled as a multi-classification problem and adopts the method of supervised learning. Moreover, the identification model can output the probability of each fault type and, subsequently, confirm the type with the highest probability as the primary fault identity. In addition, accurate and rapid fault localization after a failure can significantly reduce the fault repair time. The failed element and specific position (e.g., the localization of failed node, link, or equipment) need to be localized in the network. With the help of alarm data, the root alarm can be found through alarm correlation analysis, and the failed element can be located based on the localization information of the root cause. 

Failure diagnosis focuses on analyzing the fault after it occurs ~\cite{panayiotou2018leveraging}. This step includes confirming the type, magnitude, cause, localization, and time of the failure. Conventional failure diagnosis is generally performed by operation and maintenance personnel who perform the functions of judgment and fault auxiliary analysis. The fault auxiliary analysis function is based on the general statistical analysis method of network management data. However, it lacks the ability to comprehensively analyze a large volume of data. Therefore, intelligent diagnosis is needed to improve operation and maintenance efficiency. 

\subsection{Managed Objects in Optical Networks and Typical Failure Categories}
\begin{figure}[htbp]
\centering
\includegraphics[width=16cm]{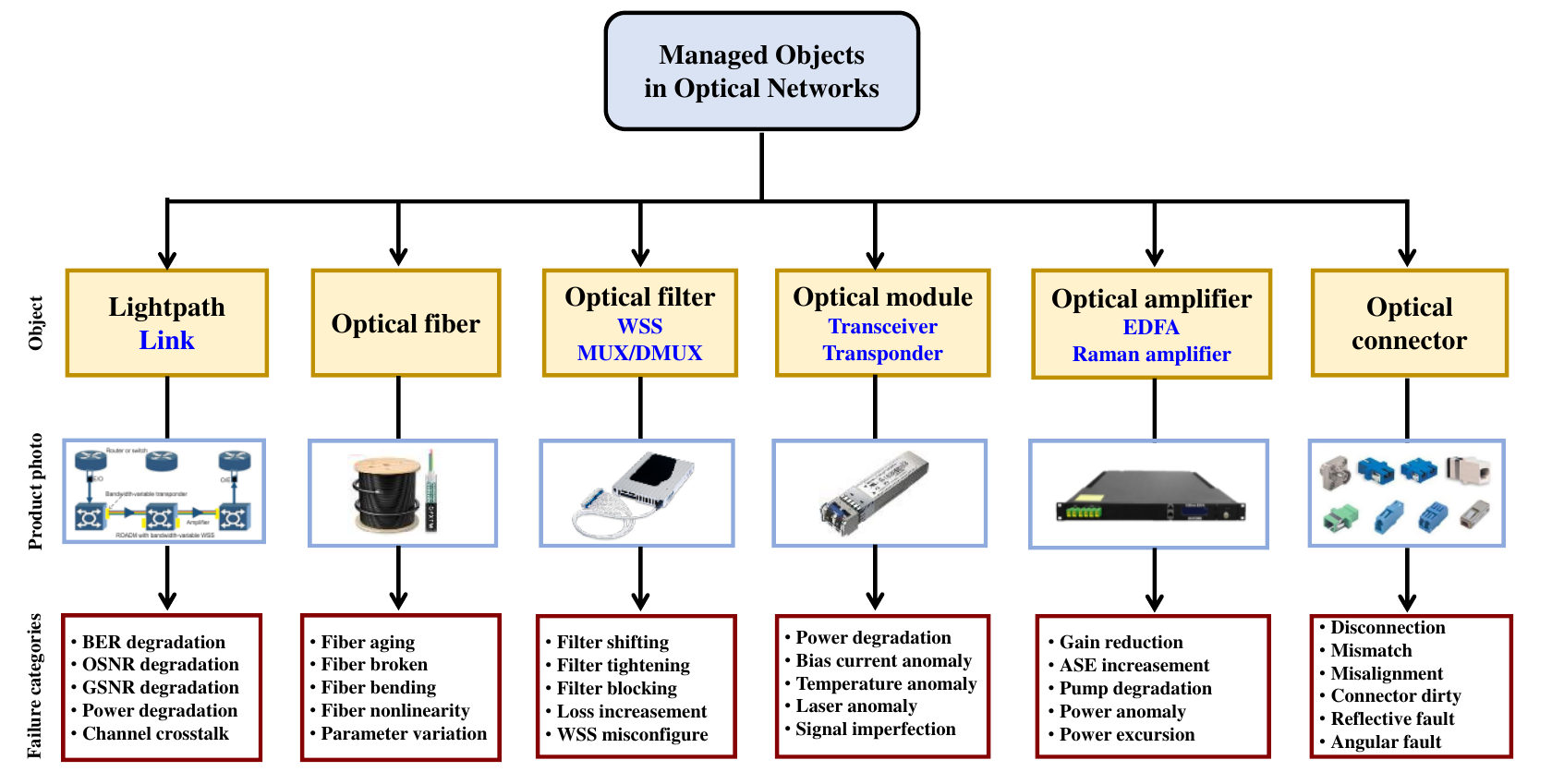}
\caption{Components of optical networks: monitored objects for failure management; the product photos for each object; the typical failures of each object.}
\label{fig2}
\end{figure}
In optical networks, failure management is implemented for some core objects, such as lightpath, optical fiber, optical filter, optical module, optical amplifier, optical connector, or optical equipment integrating these optical components, as shown in Fig. 2. Different objects suffer from different failure categories, summarized as follows:

\textbf{Lightpath}: a section of an optical network that light travels through without being modified. A lightpath is a path that carries the optical signal on a lightwave from source to destination node within optical domain. The optical signal is essentially transmitted from the transmitter, over multi-span fiber, amplifier, filter, to the detector at the destination, which is a complete optical transmission process ~\cite{barzegar2021soft}. For a lightpath, the typical failures types are bit error rate (BER) degradation, optical signal to noise ratio (OSNR) degradation, generalized SNR (GSNR) degradation, optical power drop, channel crosstalk, etc. Failure management on lightpath is a system-level management scheme concerning the overall system performance metrics ~\cite{musumeci2022domain,musumeci2020transfer}.

\textbf{Optical fiber}: a physical channel that transmits optical signal along its axis by total internal reflection. Fiber is the core component of optical networks and is the link connecting two points. In optical networks, fiber optic cables are ubiquitous and widely deployed, which are frequently damaged by wildlife, stray animals, bad weather, natural disasters, or construction work. Thus, fiber failure is the most common and major failure source. The typical fiber failures include but are not limited to fiber aging, fiber break, fiber bending, fiber nonlinearity, parameter variations (like loss increase). Most failure management tasks focus on fiber impairment estimation, diagnosis, and broken point localization ~\cite{abdelli2021reflective}.

\textbf{Optical filter}: a device with wavelength-dependent transmission or reflectance. An optical filter selectively permits the passage of a range of wavelengths, while blocking others. There are various types of filters, such as bandpass, notch, and edge filters. In optical networks, optical filters include MUX, DMUX, wavelength selective switch (WSS) enabled by array waveguide grating (AWG), or liquid crystal on silicon (LCoS). They are wavelength management components and primarily used for tasks such as wavelength division multiplexing/demultiplexing, dynamic control of channel width, wavelength selection, wavelength switch, optical equalization. Filter failures occur frequently and are generally soft failures, such as filter shifting, filter tightening, filter blocking, loss increase, WSS misconfiguration, etc ~\cite{shariati2019learning}.

\textbf{Optical module}: a hot-pluggable optical transceiver used for optical signal transmitting and receiving. Optical modules typically integrate laser, electrooptical modulator, photoelectric detector, amplifier, digital to analog converter (DAC), analog to digital converter (ADC), digital signal processing (DSP) chip, and circuit control units. Optical modules encounter several types of failure, including launch power degradation, bias current anomaly, temperature rise, laser anomaly, wavelength drift, signal performance imperfection, receiving sensitivity deterioration. As a packaged active component, failure check and troubleshooting of the optical module is always challenging and is conventionally performed by experienced engineers ~\cite{liu2022implementation}.

\textbf{Optical amplifier}: an optical repeater that amplifies optical signals directly without optical-to-electrical conversion. Optical amplifier is extremely important for long-distance optical transmission. Based on different physical mechanisms, several types of optical amplifiers exist, including erbium-doped fiber amplifier (EDFA), Raman amplifier (RA), semiconductor optical amplifier (SOA), and optical parametric amplifier. In optical networks, EDFA is the most widely applied commercial amplifier due to its high energy conversion efficiency and large gain with little crosstalk ~\cite{kruse2021edfa}. However, it also has a relatively high noise figure (NF) and frequency-dependent gain profile. In addition, RA based on stimulated Raman scattering (SRS) is another viable option especially for wideband optical transmission (C+L-band) owing to its low NF, distributed amplification, and wide gain spectrum. The RA is always combined with EDFA for hybrid amplification. The typical failure types found in optical amplifiers are gain reduction, output power anomaly, amplifier spontaneous emission (ASE) noise increasement, aging of pump lasers, and incremental power excursion.

\textbf{Optical connector}: a passive component that enables quicker connection and disconnection between two optical fibers. The connectors mechanically couple and align the cores of fibers so that light can pass from one fiber to the other one ~\cite{lefevre1993failure}. There are various types of optical connectors (LC, SC, FC, ST) and optical terminals (PC, APC, UPC). Under normal conditions, the terminals and connectors should be well matched and perfectly joined; any defects in the connectors may cause failures and extra loss, such as disconnection, mismatch, misalignment, dirt on cross-sectioning, reflective fault, angular fault.

\subsection{AI for Failure Management}
\begin{figure}[htbp]
\centering
\includegraphics[width=15cm]{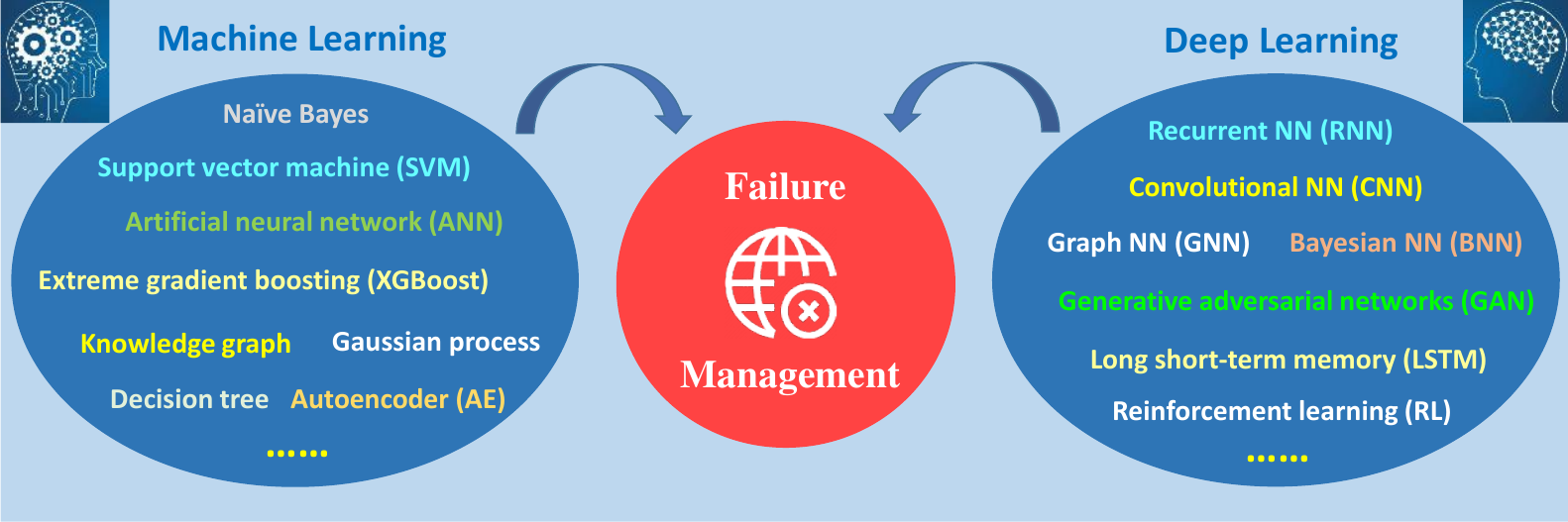}
\caption{Machine learning and deep learning are applied to failure management to solve various tasks.}
\label{fig3}
\end{figure}

In optical networks, the main objective of failure management is to guarantee above-depicted devices stable operation, proactive protection, and quick recovery. However, traditional failure management still requires complex and time-consuming human intervention. To drive the failure management towards intelligence and efficiency, techniques from AI have been widely applied to address above tasks, evolving from the early ML to recent DL, as shown in Fig. 3. ML is a main branch of AI based on the idea that systems can learn from data, identify patterns, and make decisions with minimal human intervention ~\cite{gu2020machine}. In ML families, many famous and powerful algorithms have been studied for failure management, such as support vector machine (SVM), Naïve Bayes, decision tree, artificial neural network (ANN), extreme gradient boosting (XGBoost), autoencoder, Gaussian process, etc. According to the operation function, ML algorithms can be typically divided as classification algorithm and regression algorithm, which are both supervised learning algorithms. The regression is a process to find the correlations between dependent and independent variables, which are used to predict the continuous values, such failure prediction, alarm analysis, or state parameters fitting; the classification is a process to categorize the data into different classes, which are used for failure detection, identification, or localization.

Driven by the growth of data volumes and improvement of computing power, ML have successfully evolved into DL to handle the more complex and large-scale problems with robust, adoptable, and powerful solutions ~\cite{wang2021artificial}. As the subset of ML, DL can be generally understood as deep neural network (DNN) with multiple nonlinear layers. Among DL communities, recurrent neural network (RNN), convolutional neural network (CNN), graph neural network (GNN), Bayesian neural network (BNN), generative neural network (GAN), and their variants have made distinctive contribution to pattern recognition, time series data processing, correlation analysis, and data enhancement. Furthermore, deep reinforcement learning (DRL) has made great breakthroughs in solving complicated controlling problems based on environment-aware mechanism. DL plays an important role in perception that can acquire the observation information from environment and provide the current state information, while DRL shows powerful advantages in decision-making that can sense complex system states and learn the optimal policies through repeated interactions with the environment.

\subsection{Data and Information Collected from Optical Networks for Failure Management}
\begin{figure}[htbp]
\centering
\includegraphics[width=16cm]{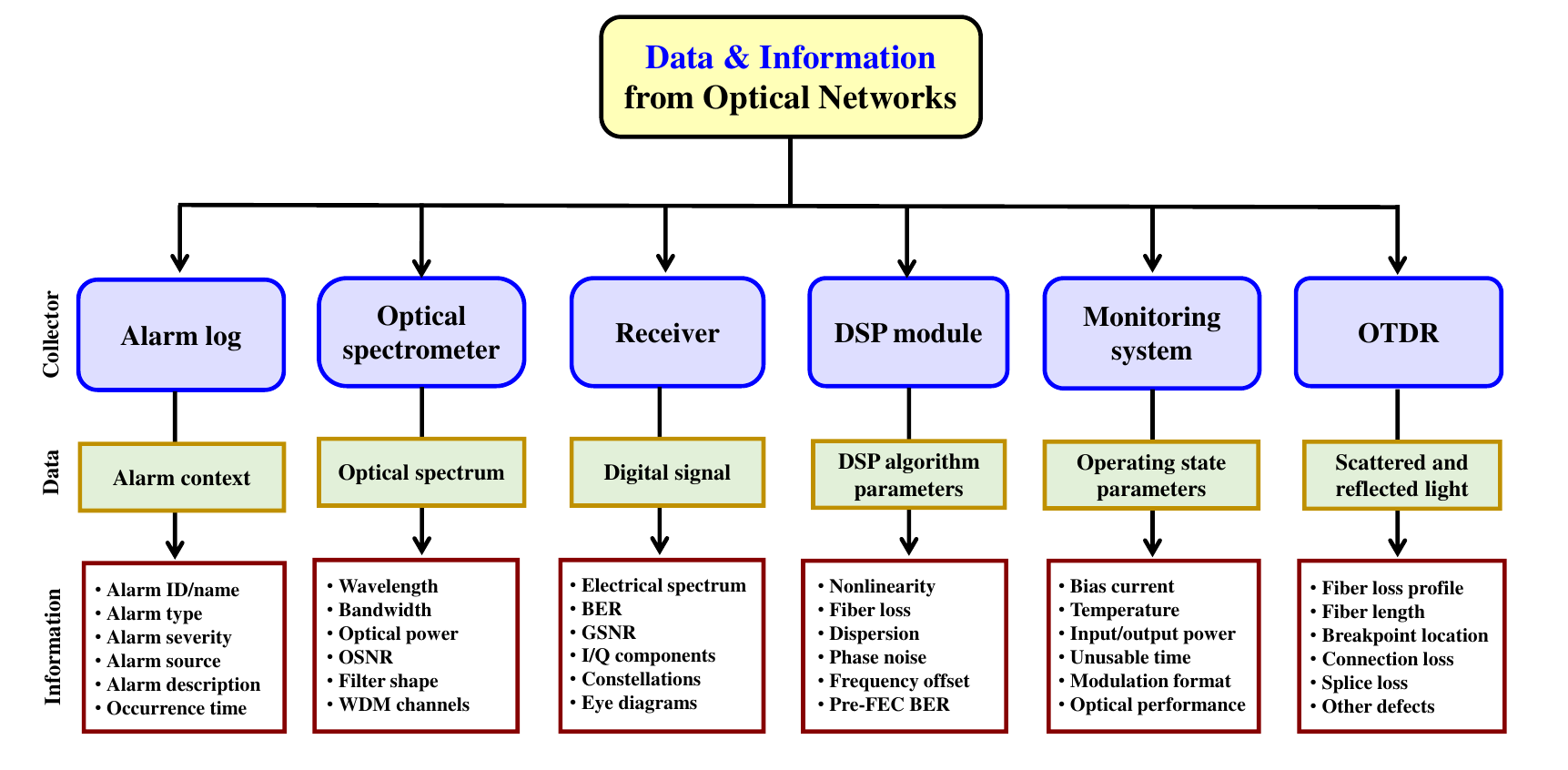}
\caption{Data collector, collected data, and extracted information from optical networks for failure management.}
\label{fig4}
\end{figure}

Basically, ML models are trained by data and learn from extracted information, that is, the more data you provide to an ML system, the faster and better that model can learn and improve. Thus, the performance of learning algorithms depends heavily on data quantity and quality. In optical networks, data can come in several forms that can be roughly divided into numerical data, time series data, text data, image data, and statistical data ~\cite{valcarenghi2021scalable,sgambelluri2021reliable}. The realization of failure management is based on the collected data pertaining to all elements and over the entire lifetime of the system. Here, we summarize data collectors in optical networks and the collected data and extracted information that are used for ML-based failure management:

\textbf{Alarm log}: as events that might require attention occur, the network management system (NMS) raises an alarm. Millions of alarms may appear in optical networks every month, containing a wealth of failure information but also including a mass of false or nuisance alarms ~\cite{stanic2010active}. Alarm logs are text data containing rich information—alarm ID/name, alarm type, alarm severity, alarm source, occurrence time, etc. For a specific alarm, the detailed illustration (definition, type, impact) is reported as context. Dealing with alarm data is important for failure management, including alarm compression, alarm correlation analysis, and alarm root cause identification.

\textbf{OSA}: optical spectrum analyzer (OSA) is an instrument used to measure the properties of optical spectrum and display the distribution of power over a specified wavelength range ~\cite{wang2019machine}. OSA is one of the most useful tools for optical performance monitoring (OPM), quality of signal (QoS) analysis, and network resource management (NRM) ~\cite{locatelli2021spectral}. Thus, optical spectral data are crucial for failure management, and lots of information can be obtained from the optical spectrum, such as channel distribution condition, optical signal parameters (wavelength, bandwidth, OSNR, power), and non-ideal filter effects. By analyzing spectral characteristics, multiple distortions can be identified, contributing to soft failure detection and prediction. 

\textbf{Receiver}: digitalized electrical signal obtained at the receiver after photoelectric detection and ADC. Digital signals in time domain carry abundant information that can comprehensively reflect the joint impairments of transmitter, fiber channel, and receiver ~\cite{tanaka2021monitoring}. After digital signal processing (DSP), the electrical spectrum can be easily obtained through Fourier transform. Meanwhile, the eye and constellation diagrams, which are the commonly-used analysis objects, are generated from the digital signal ~\cite{wang2017intelligent,wang2017modulation}. These diagrams present the amplitude and phase information of signals in all types of modulation formats. The distorted eye and constellation diagrams qualitatively present the degrees of impairments, such as nonlinear phase noise (NLPN), linear noise, I/Q imbalance, and skew effect. In addition, multiple ultimate performance indicators such as bit error rate (BER), extinction ratio (ER), error vector magnitude (EVM), and Q-factor can be calculated. 

\textbf{DSP module}: an algorithm module embedded in a digital coherent receiver is used for signal recovery, including de-skew, orthogonalization, normalization, digital equalization, timing recovery, interpolation, carrier estimation, etc. To mitigate the effect of various impairments, plenty of DSP algorithms are investigated based on the specific physical principle and insight of transmission process. Thus, by analyzing the parameters of DSP algorithms, certain impairment information such as fiber nonlinearity, loss profile, dispersion, phase noise, frequency offset, pre-FEC BER can be derived directly without other hardware ~\cite{sasai2021digital,sasai2020digital,lun2021roadm}. DSP-based failure management is a cost-effective and easily-implementable solution. 

\textbf{Monitoring system}: a supervisory system that observes the operating state and health condition of the optical network or equipment in real-time ~\cite{dong2016optical}. In general, most optical equipment include a monitoring module that can record certain parameters, such as bias current, environment temperature, input/output power, unusable time. In addition, lots of OPMs and optical channel monitors (OCMs) are deployed over the entire optical network that are used for estimation and acquisition of different performance parameters of optical signals and optical components ~\cite{wang2021optical}. The monitoring system is indispensable in ensuring reliable network operation and fault-aware proactive maintenance.

\textbf{OTDR}: an optical time domain reflectometer (OTDR) is an instrument used to characterize a fiber cable and locates events and faults along a fiber and is typically deployed in optical networks. An OTDR launches a series of optical pulses into the fiber to be measured ~\cite{shi2019event}. Various events on the fiber generate a Rayleigh back scatter that returns to OTDR. The strength of scattered or reflected light is gathered and integrated as a function of time, and plotted as a function of fiber length. OTDR provides detailed information on the localization and overall condition of fiber, connection, and splice loss, among other defects ~\cite{wu2019dynamic}. Thus, OTDR is often used to precisely detect faults in a fiber link, especially for searching for and locating the break point.

\section{Studies on Machine Learning-based Failure Management in Optical Networks}
In this section, we reviewed the studies on ML-based failure management in context of optical networks, including alarm analysis, failure prediction, failure detection, failure localization, failure identification, and failure magnitude estimation:

\subsection{Alarm Analysis}
In optical networks, alarm is the key indicator for failure management and will be reported whenever something goes wrong. In optical networks, alarm analysis includes alarm compression, trivial alarm identification, alarm prediction, alarm correlation analysis, and alarm root cause identification, as summarized in Table 1.

In recent years, some techniques explored the alarm prediction and root cause analysis in optical networks. Alarm prediction focuses on forecasting the probability of alarms in advance to make active defense. Root cause analysis concentrates on mining the correlation among multiple alarms for discovering the root cause and performing fault localization to ensure service level agreement (SLA). In ~\cite{zhao2018soon}, Y. Zhao et al. proposed an alarm prediction scheme using ANN, the input of which was the performance data and alarm type, and the output was the probability of whether the alarm would occur in the next 15 minutes. Then, they continued to study the impact of data imbalance on alarm prediction ~\cite{yan2019dirty}. To balance the data set for mean distribution, the total data volume was reduced by random deletion, and the amount of alarm data was increased by using limited Gaussian noise (LGN). The experimental results showed that data enhancement significantly improved the performance of dirty-data-based alarm prediction. Besides, a series of alarm prediction schemes were widely studied with different learning algorithms and architectures, such as random forest ~\cite{zhang2019cognitive}, GAN ~\cite{zhuang2020machine}, transfer learning ~\cite{zhang2020transfer}, and coordination between control layer AI and on-board AI ~\cite{zhao2020coordination}.

In addition to alarm prediction, alarm root cause analysis-enabled failure localization has also attracted wide attention. In ~\cite{liu2019application}, the alarm data were transmitted to network management system and analyzed by LSTM to locate the fault. The output was the possible fault position with the probability from 0 to 1 based on fuzzy theory. In ~\cite{zhao2019accurate}, an optical network fault localization method was explored using deep neural evolution network (DNEN) with large-scale alarm sets, which can make full use of the global searching ability of DNEN to perform the high-accuracy and low-latency fault localization. In ~\cite{li2020demonstration}, concept of knowledge graph (KG) was introduced to build an easy-to-understand alarm knowledge system for alarm relation reasoning, as shown in Fig. 5. Meanwhile, graph neural network (GNN) was used to recognize the root cause by inferring the relationship between alarms. Alarm KG not only visualized network alarm knowledge, but also located network faults by knowledge reasoning. However, how to build an alarm KG automatically is not an easy task. In ~\cite{zhuo2020demonstration}, an automatic construction of KG was proposed for fault localization by extracting the relationship between alarms and faults from the semi-structured data and structured data. More details of KG-based alarm analysis for fault localization can be found in ~\cite{li2021fault}.

\begin{figure}[htbp]
\centering
\includegraphics[width=15cm]{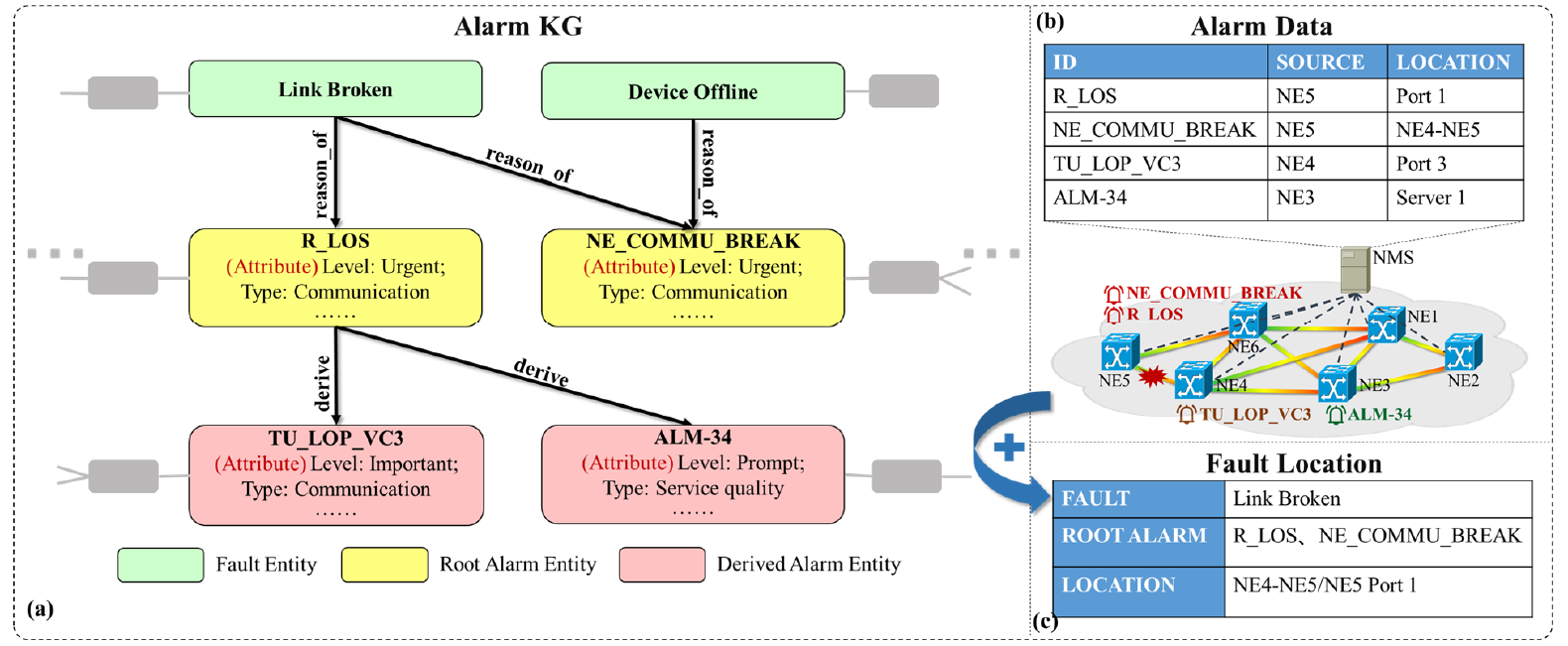}
\caption{Knowledge graph (KG)-enabled root cause analysis based on alarm data: (a) a subgraph of the alarm KG, consisting of three main types of entities (fault entities, root alarm entities, and derived alarm entities) and two types of relation edges in the graph; (b) alarm data in optical networks; (c) fault location based on alarm KGs~\cite{li2020demonstration}.}
\label{fig5}
\end{figure}

With the expansion of network scale, massive amounts of alarms come along with emergency of failures, which also brings great challenges to network management. In ~\cite{lou2018alarm, wang2019dealing}, the problems of alarm compression and alarm correlation analysis were studied. The combined K-means and ANN were used to evaluate the alarm importance quantitatively and determine alarm weights. The a-priori was modified to enhance the mining efficiency and improve the alarm compression rate, where alarm data were significantly compressed by 60$\%$ to 90$\%$. In addition, massive false or nuisance alarms distributing in different layers with different severity degrees, and thus it is difficult to identify the alarm root cause among these redundant and intricate alarms. To overcome this problem, the representation of alarm data was realized by using BERT (Bidirectional Encoder Representations from Transformers) as the pre-training model to vectorize the alarm context and using Transformer to realize alarm root cause identification ~\cite{jia2021transformer}. The alarm representation based on context vectorization could retain the multi-dimensional original information and facilitate the model training. Compared with the traditional symbol representation, the experimental results showed that the BERT could extract knowledge and read the alarm manual intelligently. Moreover, the vectorized alarm information were used to identify the root cause through transformer encoder with satisfactory accuracy.

\begin{table}[H]
\footnotesize
\centering
\caption{Summary of ML-based Alarm  Analysis}
\label{tab1}
\renewcommand\arraystretch{1.2}
\resizebox{\textwidth}{!}{%
\begin{tabular}{c|c|l|l}
\hline
Task &
  Algorithm &
  \multicolumn{1}{c}{Description} &
  \multicolumn{1}{c}{Literature} \\
\hline
\multirow{7}{*}{Alarm prediction} &
  ANN + Random forest &
  \begin{tabular}[c]{@{}l@{}}Predicting the probability of \\ an alarm occurrence in future.\end{tabular} &
  \begin{tabular}[c]{@{}l@{}}2018OE  ~\cite{zhao2018soon}\\2019OE ~\cite{yan2019dirty}\\ 2019ACP ~\cite{zhang2019cognitive}\end{tabular} \\
\cline{2-4}
 &
  GAN &
  \begin{tabular}[c]{@{}l@{}}Alarm prediction with data \\ augmentation based on GAN\end{tabular} &
  2020ICNC ~\cite{zhuang2020machine} \\
\cline{2-4}
 &
  Transfer learning &
  \begin{tabular}[c]{@{}l@{}}Concurrent alarm \\ prediction with fewer data, \\ less training time, and better efficiency.\end{tabular} &
  2020ACP ~\cite{zhang2020transfer} \\
\cline{2-4}
 &
  On-board AI &
  \begin{tabular}[c]{@{}l@{}}Supporting control layer AI and \\ on-board AI simultaneously based on SDON\end{tabular} &
  2020JOCN ~\cite{zhao2020coordination} \\
\hline
\multirow{9}{*}{Alarm root cause analysis} &
  LSTM &
  \begin{tabular}[c]{@{}l@{}}Alarm root cause \\ analysis-enabled failure localization\end{tabular} &
  2019CC ~\cite{liu2019application} \\
\cline{2-4}
 &
  \begin{tabular}[c]{@{}c@{}}DNEN\end{tabular} &
  \begin{tabular}[c]{@{}l@{}}Accurate fault localization \\ with large-scale alarm sets using DNEC\end{tabular} &
  2019OFC ~\cite{zhao2019accurate} \\
\cline{2-4}
 &
  Knowledge Graph + GNN &
  \begin{tabular}[c]{@{}l@{}}Construction of alarm \\ knowledge graphs for alarm \\ relation reasoning and fault localization\end{tabular} &
  \begin{tabular}[c]{@{}l@{}}2020OFC ~\cite{li2020demonstration}\\ 2020OFC ~\cite{zhuo2020demonstration} \\ 2021JLT ~\cite{li2021fault}\end{tabular} \\
\cline{2-4}
 &
  BERT + Transformer &
  \begin{tabular}[c]{@{}l@{}}Using BERT to vectorize \\ alarm context and using a transformer \\ to identify the alarm root cause\end{tabular} &
  2021ECOC ~\cite{jia2021transformer} \\
\hline
Alarm compression &
  K-means + ANN &
  \begin{tabular}[c]{@{}l@{}}Alarm data were compressed by\\  60\% to 90\% by evaluating the \\ alarm importance and weights\end{tabular} &
  \begin{tabular}[c]{@{}l@{}}2018OECC ~\cite{lou2018alarm}\\ 2019Access ~\cite{wang2019dealing}\end{tabular}
\\ \hline

\end{tabular}%
}
\end{table}

\subsection{Failure Prediction}
Most failure management schemes can only passively protect the optical network and minimize damage only after a failure occurs. Therefore, services are interrupted owing to the time delay of protection and recovery. Thus, failure prediction as a proactive approach is required for early-warning and proactive protection, which aims at preventing disruption in advance. Typically, failure prediction always needs to monitor the operation state and health condition of the lightpath and optical components, and then proactively switch to a backup link before failure occurs. 

\begin{figure}[!t]
\centering
\includegraphics[width=12.5cm]{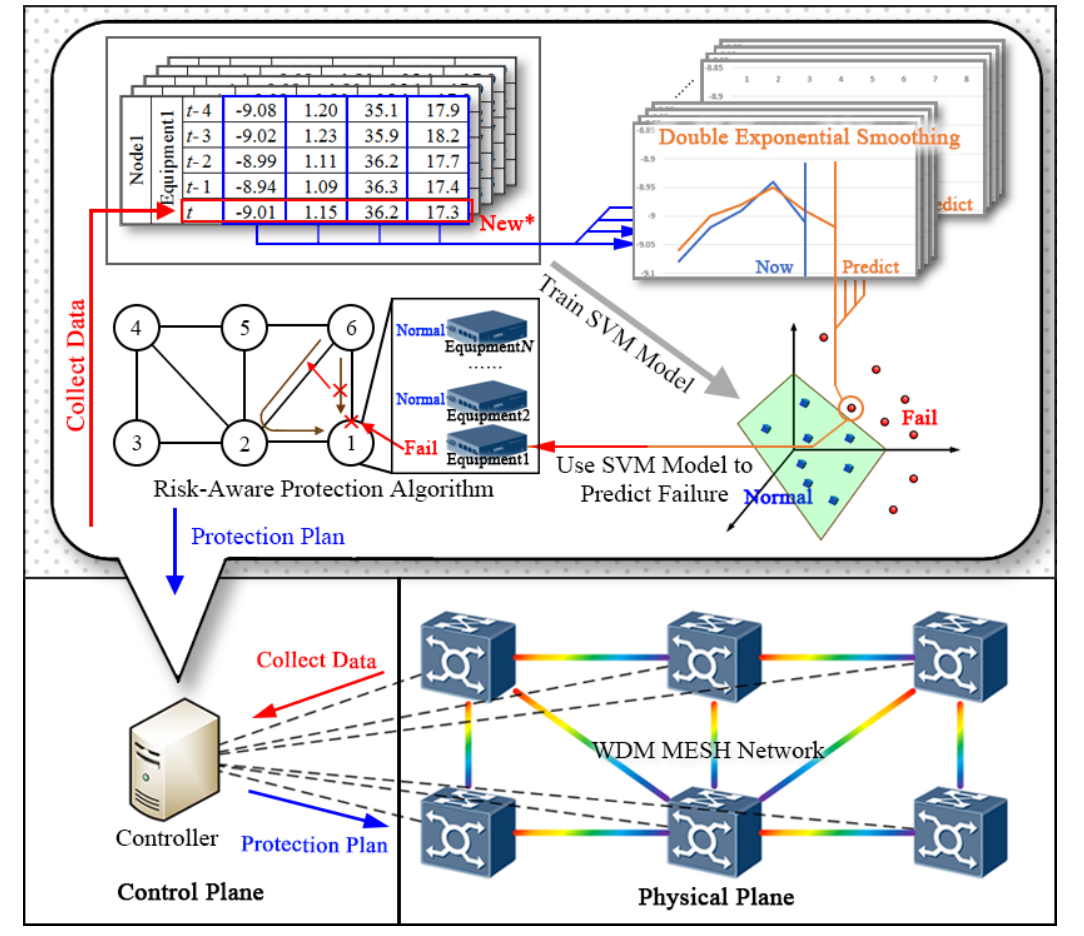}
\caption{Failure prediction of OTN equipement using SVM~\cite{wang2017failure}.}
\label{fig6}
\end{figure}

First, for the lightpath, a Gaussian process classifier was adopted to predict the failure probability of single link ~\cite{panayiotou2018leveraging} that was based on historical data extracted from the examined network using a graph-based correlation heuristic. A high overall accuracy—99$\%$ for a network load of 20 Erlangs—was achieved without any probing or other monitoring equipment. In a broad sense, the quality of transmission (QoT) estimation and bit error rate (BER) estimation could be regarded as assistive technologies for lightpath failure prediction. Estimating the lightpath QoT prior to deployment is essential for the optimized design and planning of optical networks. ML models have been extensively deployed for QoT estimation. In comparison with analytical models, ML models have exhibited superior robustness to parameter inaccuracies and lower computation complexity ~\cite{lu2021performance}. In addition to QoT, BER is the other key factor for straightforward system performance representation. In ~\cite{shariati2018autonomic}, a naïve Bayes was used for BER degradation prediction based on real-time state-of-polarization (SOP) monitoring with low complexity, minimal storage, and high speed. In ~\cite{inuzuka2020demonstration}, a framework of proactive maintenance was designed for fiber diagnosis. Here, a CNN was adopted to estimate the bend state of remote fiber; protection switching was executed as soon as fiber degradation was detected with no service disruption or bit loss.

The optical transport network (OTN) board is an important optical equipment that encapsulates the client signal in the corresponding frame format and transmits it transparently and efficiently. It primarily comprises an optical transponder unit, cross connection unit, optical multiplex/demultiplex unit, optical amplifier, and optical/electrical supervisory channel. However, the failure OTN equipment may cause massive service interruption. In connection with the OTN equipment, we proposed a series of schemes for failure management. In ~\cite{wang2017failure}, a double exponential smoothing (DSE) algorithm was used to predict the trend of operating state parameters of OTN equipment, and SVM was used to determine whether it would fail at least one day in advance, as shown in Fig. 6. This is one of the earliest studies utilizing ML for failure prediction in optical networks. In ~\cite{zhang2020temporal}, a bidirectional gated recurrent unit (BiGRU)-based adaptive failure prediction was proposed in terms of the temporal characteristic of equipment state data. In addition, principal component analysis (PCA) was used to compress data dimensionality to reduce learning complexity. Subsequently, BiGRU was used to capture the bidirectional temporal correlation between low-dimension data. Since the learning model is usually a black-box model, it is difficult to analyze its internal mechanism and determine the principal causes that induce failure. Thus, in ~\cite{zhang2021attention,wang2021uncertainty}, an attention mechanism-driven LSTM was studied for failure prediction accompanied by simultaneous possible cause identification. Besides, the above-mentioned studies only directly predict whether failure would occur, but without any uncertainty analysis. To solve this remaining issue, a Bayesian neural network (BNN) was leveraged to validate the prediction results, i.e., provide a quantified uncertainty value obtained from information entropy ~\cite{wang2021uncertainty}.

Closely related to failure prediction, several studies exist on forecasting the lifetime of optical components (primarily for lasers). This task is a more elaborate management process conducted on aging devices for its entire lifetime. Conventional aging tests are expensive, and have long test periods and low accuracy. In practice, all optical materials degrade over time, particularly in high average power or intensity optical systems. In ~\cite{smalakys2021predicting}, a Bayesian lifetime analysis approach was developed for optical materials by performing maximum a posteriori probability (MAP) estimation. Accurate and repeatable extrapolation results were obtained in S-on-1 laser-induced damage fatigue experiments. Besides, in ~\cite{abdelli2020lifetime}, an ANN-based lifetime prediction of 1550 nm DFB laser was presented to output the possible mean-time-to-failure, which improved prediction error from 3.8$ \sim $17 years to 1.12 years. Based on synthetic data and real laser datasheet including different operating conditions, this scheme effectively reduced the time and cost of aging tests, as summarized in Table 2.

\begin{table}[H]
\footnotesize
\centering
\caption{Summary of ML-based Failure Prediction}
\label{tab2}
\renewcommand\arraystretch{1.2}
\resizebox{\textwidth}{!}{%
\begin{tabular}{c|c|l|l|l}
\hline
Object &
  Algorithm &
  \multicolumn{1}{c|}{Data} &
  \multicolumn{1}{c|}{Description} &
  \multicolumn{1}{c}{Literature} \\ \hline
\multirow{6}{*}{Lightpath} &
  Gaussian process &
  \begin{tabular}[c]{@{}l@{}}Historical data: network state \\ and past failure incidents\end{tabular} &
  \begin{tabular}[c]{@{}l@{}}Predicting the failure \\ probability of a single link\end{tabular} &
  2018JOCN ~\cite{panayiotou2018leveraging} \\ \cline{2-5}
 &
  ANN vs. GN model &
  System parameters &
  \begin{tabular}[c]{@{}l@{}}Performance comparison \\ between ML and GN model for \\ QoT estimation in WDM system\end{tabular} &
  2021JOCN ~\cite{lu2021performance} \\ \cline{2-5}
 &
  Naïve Bayes &
  State of polarization (SOP) data &
  \begin{tabular}[c]{@{}l@{}}Degradation prediction \\ of pre-FEC BER based on SOP\end{tabular} &
  2018ECOC ~\cite{shariati2018autonomic} \\ \hline
Fiber &
  CNN &
  Constellation data from receiver &
  \begin{tabular}[c]{@{}l@{}}Estimating the bend \\ sate of remote fiber\end{tabular} &
  2020JLT ~\cite{inuzuka2020demonstration} \\ \hline
\multirow{7}{*}{OTN equipment} &
  DES + SVM &
  \multirow{7}{*}{\begin{tabular}[c]{@{}l@{}}Historical data: \\ operating state parameters \\ monitored from OTN equipment\end{tabular}} &
  \begin{tabular}[c]{@{}l@{}}DES for operating state \\ prediction and SVM used to \\ predict fault one day in advance\end{tabular} &
  2017OE ~\cite{wang2017failure} \\ \cline{2-2} \cline{4-5}
 &
  BiGRU + PCA &
   &
  \begin{tabular}[c]{@{}l@{}}PCA for data dimension \\ reduction and BiGRU for temporal \\ data-driven failure prognostics\end{tabular} &
  2020JOCN ~\cite{zhang2020temporal} \\ \cline{2-2} \cline{4-5}
 &
  \begin{tabular}[c]{@{}c@{}}Attention \\ mechanism-driven LSTM\end{tabular} &
   &
  \begin{tabular}[c]{@{}l@{}}Attention mechanism-driven \\ LSTM for fault prediction and \\ potential fault cause identification\end{tabular} &
  \begin{tabular}[c]{@{}l@{}}2021OFC ~\cite{zhang2021attention}\\   2021JOCN ~\cite{wang2021uncertainty}\end{tabular} \\ \hline
\multirow{3}{*}{Laser} &
  Bayesian statistics &
  \begin{tabular}[c]{@{}l@{}}Laser-induced damage \\ threshold fatigue data\end{tabular} &
  \begin{tabular}[c]{@{}l@{}}Lifetime prediction \\ by analyzing LIDT fatigue data\end{tabular} &
  2021OE ~\cite{smalakys2021predicting} \\ \cline{2-5}
 &
  ANN &
  Monitored laser parameters &
  \begin{tabular}[c]{@{}l@{}}Predicting the \\ mean-time-to-failure of a laser\end{tabular} &
  2020OFC ~\cite{abdelli2020lifetime} \\ \hline
\end{tabular}
}
\end{table}

\subsection{Failure Detection}
In contrast to failure prediction, failure detection involves triggering the alarm and examining whether a failure really occurred, when the deterioration of the monitored object reaches a certain level. In optical networks, failure detection primarily focuses on the lightpath and optical components. For failure detection in the lightpath, recent studies have adopted two different approaches: one is based on the digital coherent receiver and the other on the OSA, as summarized in Table 3.

\begin{figure}[htbp]
\centering
\includegraphics[width=8cm]{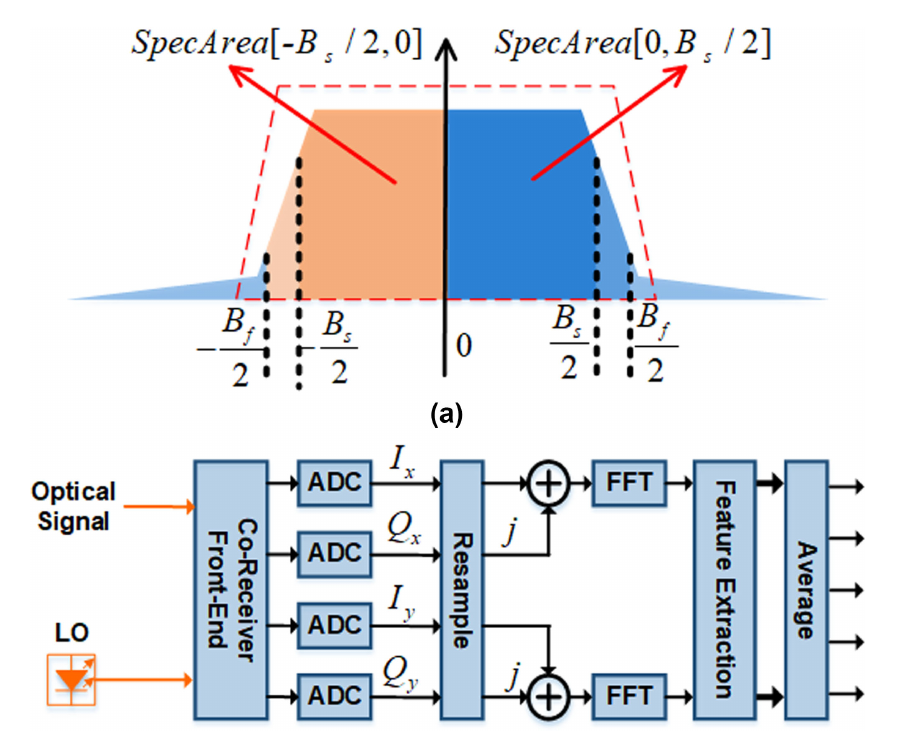}
\caption{ML-based soft failure detection exploiting digital spectrum information in coherent receiver~\cite{shu2019dual}.}
\label{fig7}
\end{figure}

In digital coherent detection systems, instead of the optical spectrum, the digital spectrum of received signals can be obtained from the coherent receiver. Moreover, with the assistance of an advanced DSP module, it identifies the rich link information and monitors multiple optical performance parameters. In ~\cite{shu2019dual,shu2019low}, a digital spectrum-assisted soft failure detection (SFD) was explored and a dual-stage SFD scheme was proposed, as shown in Fig. 7. Based on the extracted information from the digital spectrum without additional hardware, SVM was leveraged to detect soft failure caused by channel interference, filter shift, filter tightening, and increased ASE noise, and a false positive rate of 0.42$\%$ was achieved for SFD at a reasonable level of complexity. Accordingly, they further proposed an evolved digital spectrum-based SFD solution ~\cite{shu2020low}. In this study, the fast Fourier transform method was replaced with Welch’s method to reduce computational complexity by 46$\%$ and save storage cost by 99$\%$ for digital spectrum calculation. In addition, in ~\cite{lun2021gan}, a generative adversarial network-based SFD was achieved by electrical spectrum analysis that was directly obtained from the coherent receiver. In this study, by reconstructing the spectrum data from the original space to latent space, only normal data were required for training and 95$\%$ accuracy was achieved. Following this, S. Varughese et al. successively adopted SVM ~\cite{varughese2019identification} and an autoencoder ~\cite{varughese2020low} for failure detection, utilizing readily available receiver DSP features to detect multiple link failures, including inferior laser, reconfigurable optical add-drop multiplexer (ROADM) filtering, low OSNR, and adjacent-channel crosstalk.

In addition, BER degradation can represent the lightpath state intuitively. Thus, BER detection is another important tool for failure detection. In ~\cite{vela2017ber}, a BER anomaly detection algorithm was proposed to detect excessive BER within network nodes. It could trigger the affected service re-routing and minimize the effect of failure. Moreover, in ~\cite{shahkarami2018machine}, several ML algorithms were studied for SFD based on continuous monitoring BER values, called BER anomaly detection. Comparing binary-SVM, random forest (RF), multiclass SVM, and single-layer neural network, the tradeoff between accuracy and complexity was explored, and accuracy above 98$\%$ was consistently obtained.

For OSA-based failure detection methods without photoelectric conversion, all the analyzing and processing are directly executed in the optical domain. In optical networks, most monitors are based on optical spectrum measurement and lots of information can be obtained from optical spectral data, including power, OSNR, bandwidth, wavelength, spectral shape, and filtering effect. Learning from the optical spectrum, L. Velasco et al. proposed a series of schemes to detect filter-related failures, i.e., filter shift and filter tight ~\cite{shariati2018monitoring,velasco2018learning,shariati2019learning}. With the help of ML algorithms, the asymmetrical and rounded-edge features of abnormal spectra could be distinguished from normal spectra. These schemes took the advantages of ML’s powerful recognition ability to reduce the requirement on instruments, allowing coarse-resolution and cost-effective OSAs.

Furthermore, for optical components, fiber is the primary research object in terms of failure detection. In ~\cite{boitier2017proactive}, a proactive fiber damage detection scheme was realized through DSP in a coherent receiver. Learning from the features of monitored SOP, a naïve Bayes classifier was used to detect multiple fiber damages, including bending, shaking, small hit, and up and down events, with 95$\%$ reliability. In ~\cite{abdelli2021reflective}, LSTM was applied for multitask learning to detect fiber reflective faults generally occurring in connectors or mechanical splices. Learning from the noisy data collected by OTDR and a sequence of signal power levels, LSTM detected reflective events with 93$\%$ accuracy, outperforming the conventional OTDR analysis technique.

In addition to fiber, other optical components were also studied with respect to failure detection, including OTN equipment, optical modules, and laser. In ~\cite{liu2021semi}, OTN failure detection was performed with extremely imbalanced data using an autoencoder. As generally known, the monitored data collected from practical networks always present imbalanced characteristics. Thus, the failure data were relatively rare, which was a big challenge for training ML algorithm. The autoencoder could map the imbalanced data from original to latent space, where the features of normal and abnormal data could be easily distinguished, thereby significantly reducing the its dependence on failure data (\textless3$\%$). In ~\cite{rafique2017cognitive}, the concept of cognitive fault management was introduced and an initial framework was designed. By monitoring the received optical power levels, ML was employed for autonomous failure detection. Compared with conventional fixed threshold-triggered methods, it achieved superior performance in terms of accuracy and response time. In ~\cite{abdelli2019machine}, laser degradation detection was studied to enhance laser reliability. Learning from synthetic historical data, LSTM was adopted to predict its remaining working life and recognize several laser degradation modes.

Moreover, more and more researchers focus on cognitive failure management. Cognitive failure management integrates cognitive activities (such as decision-making, identification, etc.) into failure management by observing the network state, and uses the acquired knowledge to take action. In cognitive failure detection, a self-taught anomaly detection with hybrid unsupervised and supervised ML was proposed, which used unsupervised density-based clustering algorithms to learn abnormal network behavior ~\cite{chen2019self}. It eliminated the dependence on a priori knowledge of abnormal network behavior and could potentially detect unforeseen anomalies. In ~\cite{chen2022on}, a hybrid ML method was proposed for collaborative failure management in multi-domain optical networks. By analyzing the data patterns, online anomaly detection and soft failure localization were realized. The use of federated learning architecture enabled multi-domain data to effectively share knowledge in the case of high imbalance, and achieved a soft failure detection rate higher than 90$\%$. In ~\cite{furdek2021optical}, a functional module called Security Operation Center (SOC) was put forward from the perspective of optical network failure management cognition and automatic security management. The experimental results showed that dimensionality reduction and unsupervised learning techniques improved the accuracy of attack detection and reduced the run time.

\begin{table}[H]
\footnotesize
\centering
\caption{Summary of ML-based Failure Detection}
\label{tab3}
\renewcommand\arraystretch{1.2}
\resizebox{\textwidth}{!}{%
\begin{tabular}{c|c|l|l|l}
\hline
Object &
  Algorithm &
  \multicolumn{1}{c|}{Data} &
  \multicolumn{1}{c|}{Description} &
  \multicolumn{1}{c}{Literature} \\
\hline
\multirow{8}{*}{Lightpath} &
  \begin{tabular}[c]{@{}c@{}}Gaussian \\ distribution\\   OC-SVM\end{tabular} &
  \begin{tabular}[c]{@{}l@{}}Digital spectrum from receiver\end{tabular} &
  \begin{tabular}[c]{@{}l@{}}Low-complexity dual-stage \\ soft-failure detection\end{tabular} &
  \begin{tabular}[c]{@{}l@{}}2020JLT   ~\cite{shu2019dual}    \\ 2020OE ~\cite{shu2020low}\end{tabular} \\
\cline{2-5}
 &
  GAN &
  \begin{tabular}[c]{@{}l@{}}Electric spectrum from receiver\end{tabular} &
  \begin{tabular}[c]{@{}l@{}}Soft-failure detection \\ for long-haul transmission \\ systems with only normal data\end{tabular} &
  2021OFC ~\cite{lun2021gan} \\
\cline{2-5}
 &
  \begin{tabular}[c]{@{}c@{}}One-class SVM\\    Autoencoder\end{tabular} &
  \begin{tabular}[c]{@{}l@{}}Available adaptive filter \\ coefficient in DSP\end{tabular} &
  \begin{tabular}[c]{@{}l@{}}Using readily available \\ receiver DSP features to \\ detect multiple link failures\end{tabular} &
  \begin{tabular}[c]{@{}l@{}}2019OFC ~\cite{varughese2019identification}\\  2020OFC ~\cite{varughese2020low}\end{tabular} \\
\cline{2-5}
 &
  \begin{tabular}[c]{@{}c@{}}Binary-SVM\\  Finite state machine\end{tabular} &
  \begin{tabular}[c]{@{}l@{}}BER traces monitored \\ from receiver\end{tabular} &
  \begin{tabular}[c]{@{}l@{}}Anomaly detection was realized \\ by monitoring continuous BER values\end{tabular} &
  \begin{tabular}[c]{@{}l@{}}2018OFC ~\cite{shahkarami2018machine}\\ 2017JLT ~\cite{vela2017ber}\end{tabular} \\
\hline
Filter &
  \begin{tabular}[c]{@{}c@{}}Decision tree \\  SVM\end{tabular} &
  Optical spectrum from OSA &
  \begin{tabular}[c]{@{}l@{}}Optical spectrum-based \\ filter-related failures \\ (filter shift and filter tightening)\end{tabular} &
  \begin{tabular}[c]{@{}l@{}}2018OFC ~\cite{velasco2018learning}\\  2019JLT ~\cite{shariati2019learning}\end{tabular} \\
\hline
\multirow{5}{*}{Fiber} &
  Naïve Bayes &
  \begin{tabular}[c]{@{}l@{}}SOP data obtained \\ from the receiver\end{tabular} &
  \begin{tabular}[c]{@{}l@{}}Multiple fiber damages detection \\ using monitored SOP features\end{tabular} &
  2017ECOC ~\cite{boitier2017proactive} \\
\cline{2-5} 
 &
  LSTM &
  Noisy data collected by OTDR &
  \begin{tabular}[c]{@{}l@{}}Reflective fault detection based \\ on monitored data obtained by OTDR\end{tabular} &
  2021JOCN ~\cite{abdelli2021reflective} \\
\cline{2-5}
 &
  CNN &
  OPM data and eye-diagram &
  \begin{tabular}[c]{@{}l@{}}Eavesdropping detection \\ based on eye diagram and OPM data\end{tabular} &
  2022OFT ~\cite{song2022experimental} \\
\hline
OTN equipment &
  Autoencoder &
  \begin{tabular}[c]{@{}l@{}}Historical performance \\ monitoring data\end{tabular} &
  \begin{tabular}[c]{@{}l@{}}Anomaly detection with \\ extremely imbalanced data\end{tabular} &
  2021OFC ~\cite{liu2021semi} \\
\hline
Optical module &
  ANN &
  Received optical power levels &
  \begin{tabular}[c]{@{}l@{}}Fault detection based on isolated \\ optical power level abnormalities \\ across various network nodes\end{tabular} &
  2017JLT ~\cite{rafique2017cognitive} \\
\hline
Laser &
  LSTM &
  Synthetic historical data &
  \begin{tabular}[c]{@{}l@{}}Several laser degradation \\ modes detection\end{tabular} &
  2019ICTON ~\cite{abdelli2019machine}\\ \hline
\end{tabular}
}
\end{table}

\subsection{Failure localization}

\begin{figure}[htbp]
\centering
\includegraphics[width=15cm]{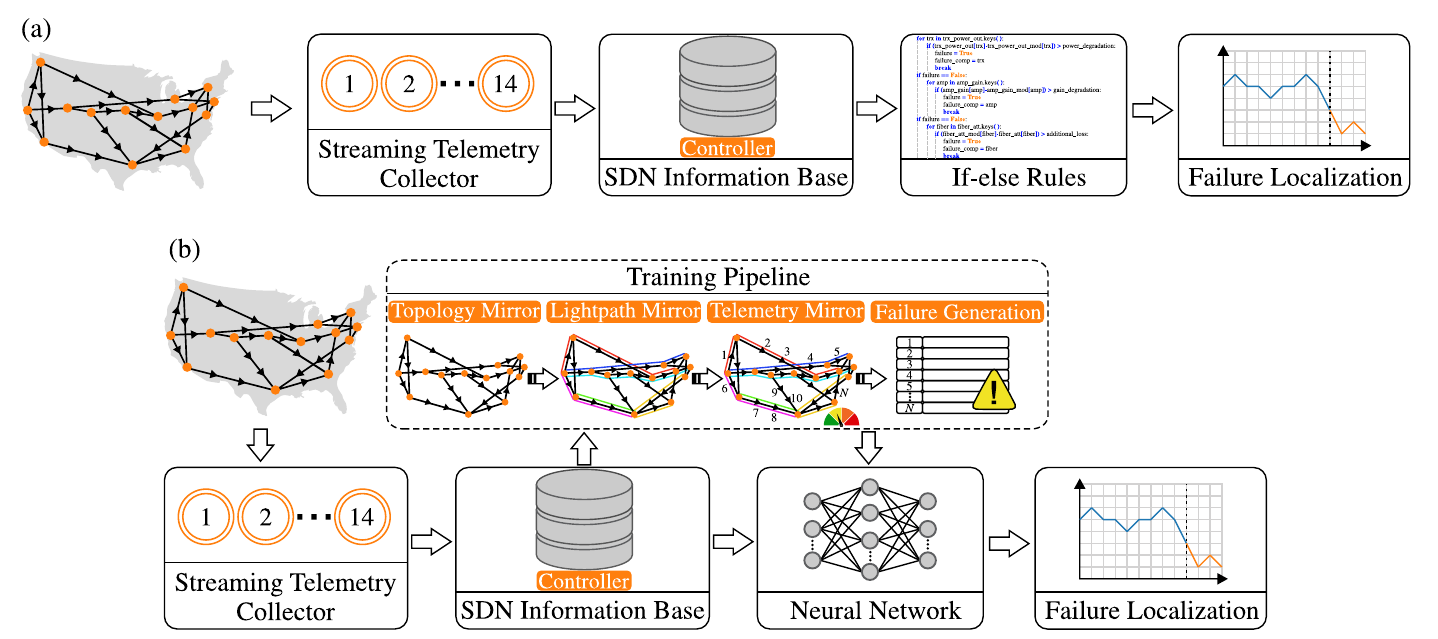}
\caption{ML-based network-wide soft-failure localization frameworks with telemetry~\cite{mayer2021machine}.}
\label{fig8}
\end{figure}
After failure detection, immediate failure localization is of paramount importance for failure recovery. In ~\cite{lun2020anomaly}, the localization of irregular WSS was inferred by power spectrum density (PSD) when filter shift occurred in different localizations of the fiber link. When failure occurred in different localizations, the PSD presented different distortion degrees, and the frequency response of the adaptive filter would converge to corresponding states that was used to spot the irregular WSS. In ~\cite{mayer2020soft,mayer2021machine}, a network-wide soft failure localization framework was designed and an ANN-based approach was proposed with SDN streaming telemetry, where training data were collected from telemetry and stored in a time-series database, as shown in Fig. 8. Multiple cases, including amplifier failure localization, fiber failure localization, and transponder failure localization, were comprehensively studied in two different topologies.

In optical networks, QoT representing system performance can be degraded by several effects. Supported by GNPy tool, QoT can be easily calculated and observed in real time. In ~\cite{barzegar2020soft}, a soft failure localization was realized by analyzing QoT-related parameters. When failure occurred in WSSs or EDFAs, the value of GSNR would decrease owing to noise figure increase, launch power decrease, fiber cut, or wavelength shift of WSSs. Based on the evolution of GSNR over time, the cause can be identified and failure can be located accordingly. 

Based on the optical spectrum, two failure localization methods were proposed to figure out WSS misconfigurations: one used ROADM with OSA and the other used an optical supervisory channel ~\cite{gifre2018experimental}. In ~\cite{vela2018soft}, relying on specifically designed optical testing channel (OTC) modules and widely deployed OSAs, the optical parameters were retrieved as training data of ML in the network controller. During commissioning testing and lightpath operation, soft failures caused by laser and filter could be identified and localized with an accuracy above 90$\%$. The accuracy of this method was further enhanced to cope with the filter cascading problem in ~\cite{velasco2018learning,shariati2019learning}, where the captured OSNR values from the optical spectrum were used to reveal the location of failure among nodes 1-4.

In ~\cite{christodoulopoulos2016exploiting}, a network Kriging (NK)-based scheme was explored for failure localization using the information from lightpaths. As a mathematical framework, NK was used to correlate physical parameters with failure elements to unambiguously localize the fiber link. In ~\cite{panayiotou2018leveraging}, a GP-based link failure localization method was proposed for service providers to reduce the network cost without using lightpath probing or other monitoring equipment. The proposed method comprised two phases: failure detection over the entire network topology. and calculation of the failure probability of each link to spot the suspected link, in the event of failure. In general, failure localization can be divided into network-level and lightpath-level cases. Optical network is composed of multiple optical links and optical nodes. Therefore, network-level failure localization includes not only all the links but also nodes, consisting of multiple lightpaths. While lightpath-level failure localization only involves the components (like transceiver, fiber link, amplifier, and filter) belonging to its own lightpath.

For fiber defect localization, OTDR is the most useful tool. Combining LSTM with OTDR, a reflective fiber fault localization method was studied to detect the reflective events, including fiber mismatch, fiber breaks, angular fault, dirt on connector, and micro-bends ~\cite{abdelli2021reflective}. This combination of methods achieved a higher detection probability and localization accuracy. In addition to OTDR-based methods, another interesting approach without other hardware or extra DSP modules was proposed for fiber degradation localization, where the loss profile and passband narrowing could be obtained in receiver-side DSP using the classic digital backpropagation (DBP) algorithm ~\cite{sasai2020digital}. This approach was based on the learning of backpropagation with complex-valued finite impulse response (FIR) filters to depict the loss conditions over the entire transmission link, which could be used for fiber anomaly detection and degradation localization. Subsequently, T. Sasai et al. extended this method to monitor more fiber parameters, longitudinal loss and dispersion profiles along a multi-span link, which was useful for accurate and convenient fiber degradation monitoring and failure localization ~\cite{sasai2021physics}. To save space, these works about failure localization are summarized in Table 4 with the following failure diagnosis together.

\subsection{Failure Identification and Failure Magnitude Estimation}
\begin{figure}[htbp]
\centering
\includegraphics[width=15cm]{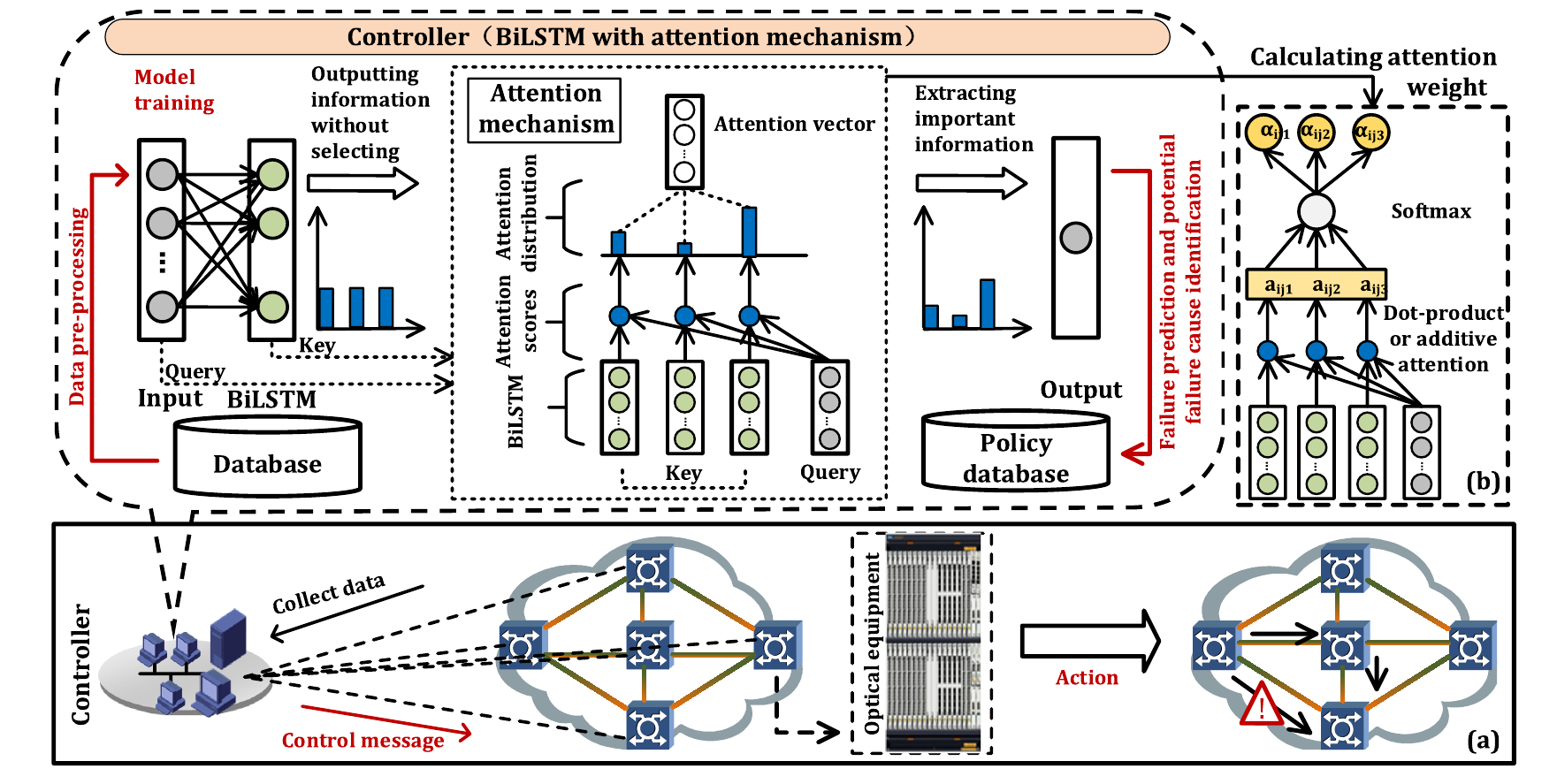}
\caption{Potential fault cause identification using attention mechanism~\cite{zhang2022potential}.}
\label{fig9}
\end{figure}

If failure really occurs, failure identification and failure magnitude estimation will be performed, which is crucial for troubleshooting and failure maintenance. Failure identification mainly includes failure type identification and failure cause identification. Recently, various failure identification schemes were explored in optical networks, as summarized in Table 4.

In general, failure identification has always been simultaneously realized with failure detection in most studies. In electric spectrum-based failure detection schemes ~\cite{shu2019dual,shu2020low}, in the event of failure, the causes could be identified by ML algorithms (SVM, CNN, or ANN). Learning from digital spectrum information in the coherent receiver, multiple impairment deteriorations could be recognized by ML-embedded DSP, including filter shift, filter tightening, filter-cascaded effect, increased ASE noise, nonlinearity, channel crosstalk, and fiber length fluctuation. For the faults caused by multiple soft failure causes, the cause identification of optical link soft failure was realized based on power spectrum density ~\cite{lun2020softfailure}. Especially, if multiple soft failure causes coexist, based on the probability information of output causes of CNN output layer, the proposed scheme can identify the main soft failure cause. In optical spectrum-based methods ~\cite{velasco2018learning,shariati2019learning}, optical spectral data were directly collected from OSA to train ML algorithms for failure identification, where filter-related effects, laser shift, and OSNR degradation could be effectively identified. Apart from spectral methods, BER monitoring was employed to anticipate connection disruption. Considering pre-FEC BER and received power together, several potential failure elements were analyzed in the centralized network controller, including signal overlap, tight filtering, gradual drift, and cyclic drift ~\cite{vela2017ber}. Based on DSP in the receiver, S. Varughese et al. employed the readily available adaptive filter coefficient to identify impairments of low OSNR, nonlinearity, inferior lasers, faulty ROADM, and inter-channel interference using SVM ~\cite{varughese2019identification} and an autoencoder ~\cite{varughese2020low}, respectively.

The above-mentioned schemes primarily identified the possible failure types; additionally, certain studies attempted to determine which specific elements induced failures in the optical network infrastructure, i.e., failure cause identification. In ML algorithms, NNs are the most widely used option for failure management. However, the black-box characteristic makes it difficult to interpret its operating principle and analyze feature importance in failure, which causes serious difficulties for the operator. To overcome this issue, XGBoost was introduced to conduct failure cause identification of OTN boards in optical networks ~\cite{zhang2020interpretable}. In contrast to NNs, XGBoost is an integrated model comprising several base learners that make XGBoost interpretable and decomposable. By analyzing the internal structure of XGBoost, five features were revealed as the most possible failure causes, and, with the assistance of SHAP, these potential causes were ranked in turn by feature importance and correlation ~\cite{zhang2021cause}. In addition, to overcome the interpretability of NN, the attention mechanism technique was introduced to drive DL, enabling the model to dynamically focus on certain input features~\cite{zhang2021attention}. With the help of attention mechanism, LSTM not only detected failures accurately, but identified failure causes, as shown in Fig. 9. According to the attention weight distribution, the two highest-relevance features for equipment failure were confirmed: the average value of input optical power and the minimum value of the environmental temperature. Thus, relevant information for failure diagnosis and maintenance was obtained ~\cite{zhang2022potential}.

In failure analysis, failure magnitude estimation is another important task required to quantify specific failure effects and understand failure severity. The failure magnitude is defined as the difference between the ideal and actual value of the estimated object. In ~\cite{wang2019machine,vela2018soft}, a laser drift estimator, filter shift estimator, and filter tightening estimator were simultaneously deployed to measure different magnitudes. In ~\cite{shu2020low}, mutual information between features and failure magnitude were obtained by a digital spectrum-based solution. Compared with real failure magnitudes, the mean squared errors (MSEs) of estimated failure magnitudes for five soft failures were all \textless0.4. In ~\cite{vela2018soft}, ANN and Gaussian process regression (GPR) were combined to estimate anomaly values of ROADM through extracting information of PSD and equalizer taps in receiver, contributing to network reconfiguration and equipment adjustment for recovery. Failure magnitude estimation could provide the specific deviation values from the normal condition, which is the detailed reference information for network reconfiguration and equipment adjustment for recovery.

\begin{table}[h]
\footnotesize
\centering
\caption{Summary of failure localization, failure identification, and failure magnitude estimation}
\label{tab4}
\renewcommand\arraystretch{1.2}
\resizebox{\textwidth}{!}{%
\begin{tabular}{c|c|l|l|l}
\hline
Task &
  Algorithm &
  \multicolumn{1}{c|}{Data} &
  \multicolumn{1}{c|}{Description} &
  \multicolumn{1}{c}{Literature} \\
\hline
\multirow{11}{*}{Failure localization} &
  ANN &
  \begin{tabular}[c]{@{}l@{}}PSD of received signals and tap \\ coefficients of adaptive filter\end{tabular} &
  Localizing the irregular WSS &
  2020OFC ~\cite{lun2020anomaly} \\
\cline{2-5}
 &
  ANN &
  Data from streaming telemetry &
  \begin{tabular}[c]{@{}l@{}}The locations of failed \\ amplifier, fiber, transponder \\ were spotted in two topologies\end{tabular} & \begin{tabular}[c]{@{}l@{}}2020ECOC~\cite{mayer2020soft}\\ 2021JOCN ~\cite{mayer2021machine} \end{tabular} \\
\cline{2-5}
 &
  GN model &
  GSNR value &
  Localizing the soft-failed WSS or EDFA &
  2020OFC ~\cite{barzegar2020soft} \\
\cline{2-5}
 &
  Gaussian process &
  \begin{tabular}[c]{@{}l@{}}Historical data: network \\ state and past failure incidents\end{tabular} &
  Spotting the suspected link &
  2018JOCN ~\cite{panayiotou2018leveraging} \\
\cline{2-5}
 &
  LSTM &
  Reflective data from OTDR &
  \begin{tabular}[c]{@{}l@{}}Fiber defect localization by \\ detecting reflective events\end{tabular} &
  2021JOCN ~\cite{abdelli2021reflective} \\
\cline{2-5}
 &
  DBP algorithm &
  \begin{tabular}[c]{@{}l@{}}Coefficients of \\ backpropagation with FIR filters\end{tabular} &
  \begin{tabular}[c]{@{}l@{}}Fiber degradation \\ localization based on estimated \\ loss and dispersion profile\end{tabular} &
  2020ECOC ~\cite{sasai2020digital} \\
\hline
\multirow{11}{*}{Failure identification} &
  \begin{tabular}[c]{@{}c@{}}SVM\\ CNN\\ ANN\end{tabular} &
  Electric spectrum from receiver &
  \begin{tabular}[c]{@{}l@{}}Filter effects, increased \\ ASE noise, nonlinearity \\ channel crosstalk, and fiber \\ length fluctuation were identified.\end{tabular} &
  \begin{tabular}[c]{@{}l@{}}2020JLT ~\cite{shu2019dual} \\ 2020OE ~\cite{shu2020low}\end{tabular} \\
\cline{2-5}
 &
  SVM &
  Optical spectrum from OSA &
  \begin{tabular}[c]{@{}l@{}}Filter effects, laser shift, \\ and OSNR degradation were identified\end{tabular} &
  2019JLT ~\cite{shariati2019learning} \\

\cline{2-5}
 &
  \begin{tabular}[c]{@{}c@{}}SVM\\ Autoencoder\end{tabular} &
  Adaptive filter coefficient in DSP &
  \begin{tabular}[c]{@{}l@{}}Nonlinearity, inferior laser, \\ faulty ROADM were identified\end{tabular} &
  \begin{tabular}[c]{@{}l@{}}2019OFC ~\cite{varughese2019identification} \\ 2020OFC ~\cite{varughese2020low}\end{tabular} \\
\cline{2-5}
 &
  CNN &
  Power spectrum density &
  \begin{tabular}[c]{@{}l@{}}Cause identification of soft failure  \\ based on power spectrum density\end{tabular} &
  2020JLT ~\cite{lun2020softfailure} \\
 \cline{2-5}
 &
  XGBoost &
  \begin{tabular}[c]{@{}l@{}}Historical data: operating \\ state parameters of equipment\end{tabular} &
  \begin{tabular}[c]{@{}l@{}}Five potential causes were \\ revealed in turn by analyzing \\ the internal structure of XGBoost\end{tabular} &
  \begin{tabular}[c]{@{}l@{}}2020OFC ~\cite{zhang2020interpretable} \\ 2021OE ~\cite{zhang2021cause}\end{tabular} \\
\cline{2-5}
 &
  \begin{tabular}[c]{@{}c@{}}Attention\\ mechanism-driven LSTM\end{tabular} &
  \begin{tabular}[c]{@{}l@{}}Historical data: operating \\ state parameters of equipment\end{tabular} &
  \begin{tabular}[c]{@{}l@{}}The highest-relevance failure \\ causes were confirmed according \\ to attention weight distribution\end{tabular} &
  \begin{tabular}[c]{@{}l@{}}2021OFC ~\cite{zhang2021attention}\\ 2022JOCN ~\cite{zhang2022potential}\end{tabular} \\
\hline
\multirow{4}{*}{Failure magnitude estimation} &
  SVM &
  Optical spectrum from OSA &
  \begin{tabular}[c]{@{}l@{}}The magnitude of laser and \\ filter drift were estimated\end{tabular} &
  \begin{tabular}[c]{@{}l@{}}2019Access ~\cite{wang2019machine}\\ 2018JOCN ~\cite{vela2018soft}\end{tabular} \\
\cline{2-5}
 &
  Welch's method &
  Digital spectrum from receiver &
  \begin{tabular}[c]{@{}l@{}}MSEs of five estimated \\ failure magnitudes were \textless{}0.4\end{tabular} &
  2020OE ~\cite{shu2020low} \\
\cline{2-5}
 &
  ANN+GPR &
  PSD and equalizer taps from receiver &
  Anomaly values of ROADM were estimated &
  2021JLT ~\cite{lun2021roadm} \\ \hline
\end{tabular}
}
\end{table}

\section{Future Research Directions for ML-based Failure Management}

In the future, several possibilities exist for the further enhancement of management ability in failure management of optical networks. First, the quality and amount of data are still the most significant issue for ML training and model building. However, in the practical networks, the failure data are rare and difficult to acquire, which results in an extremely imbalanced distribution between normal and abnormal data. Meanwhile, the raw data are always collected without labels and data annotation requires a lot of time and human labor. Thus, the challenges of small dataset and data imbalance should be solved in future studies, where the ML algorithms could learn from the limited and imbalanced failure data. 

In addition, network telemetry is being rapidly developed to enable on-demand streaming of real-time monitoring parameters. Telemetry would enhance the capability of online data collection, and thereby failure management in the future should be reinforced by telemetry. Meanwhile, telemetry enables massive data collection in real time from several optical devices provided by various services, related to different failure management tasks. However, most of existing methods can only handle single tasks. Thus, multiple tasks are mutually isolated, leading to a lack of correlation analysis. Thus, multitask learning should be further studied to develop multitasking, and correlation analysis should be conducted to empower collaboration and interaction among the tasks. 

\begin{figure}[H]
\centering
\includegraphics[width=16cm]{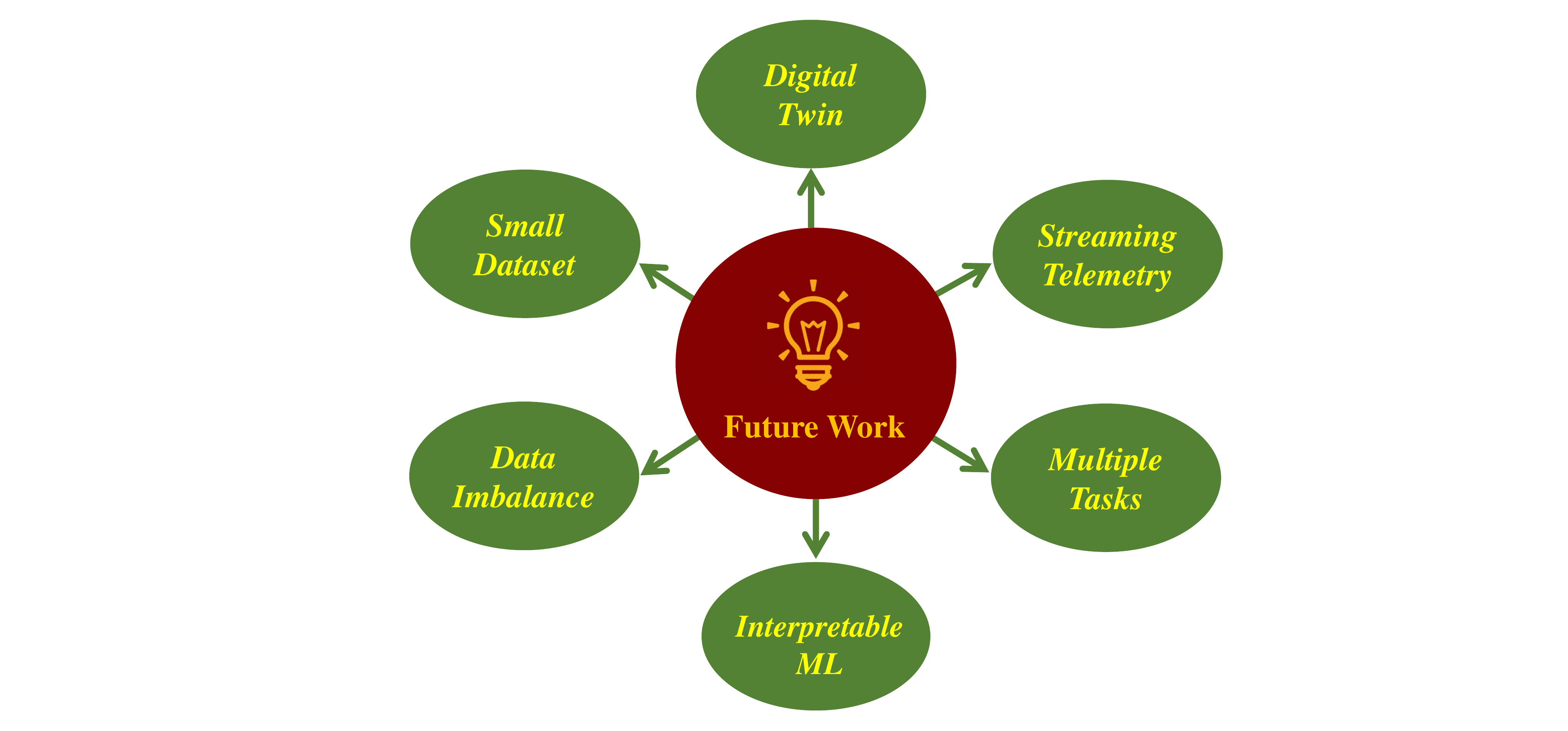}
\caption{Future work for failure management in optical networks: ML algorithms learn from small dataset and imbalanced data; telemetry assisted real-time data collection; multitask learning and correlation analysis for multiple failure tasks; digital twin for life-cycle health monitoring; physics-informed machine learning for interpretable algorithms instead of pure data-driven ML}
\label{fig10}
\end{figure}

Finally, we need to focus on the emerging techniques that have the potential to be applied in failure management. Digital twin is a powerful tool that can enable the construction of the mirror model of physical objects and monitor them in digital space, which has been introduced to optical communications to perform the life-cycle health management. Furthermore, previous ML algorithms used in failure management were based on data-driven modeling without other prior knowledge. In addition, recently, physics-informed machine learning has been attracting wide attention from various areas, because it combines the benefits of machine learning and physical principles instead of functioning according to a purely data-driven approach~\cite{jiang2021solving, jiang2021physics}. In optical communications, lots of physical knowledge have been explored and could provide helpful information and insightful analysis for failure management. This is another promising research direction for failure management from the perspective of ML innovation. Accordingly, the interpretability of ML is another promising and important research direction for optical communications~\cite{fan2020advancing}, and equally important for failure management. The interpretable ML could not only implement the basic function but also provide the insight information for failure analysis and diagnosis, which has started to appear very recently in~\cite{karandin2022if}.


\Acknowledgements{This work was supported in part by National Key R\&D Program of China (2019YFB1803502), by National Natural Science Foundation of China (No. 62171053, 61975020).}





\end{document}